\documentstyle{article}

\input epsf

\begin{document}

\title{Calculation of the Phase Behavior of Lipids 
}

\author{M.\ M\"{u}ller{${}^*$} and M.\ Schick
\\
{\small Department of Physics, Box 351560, University of Washington} \\
{\small Seattle, Washington 98195-1560} \\
{{${}^*$}\small present address: Institut f{\"u}r Physik,}\\
{\small Johannes Gutenberg Universit{\"a}t, D 55099 Mainz, Germany}
}
\date{\today}
\maketitle

\begin{abstract}
The self-assembly of monoacyl lipids in solution is studied
employing a model  in which the lipid's hydrocarbon tail is
described within the Rotational Isomeric State framework and is attached
to a simple hydrophilic head. Mean-field theory is employed, and the
necessary partition function of a single lipid is obtained via a partial
enumeration over a large sample of molecular conformations. The
influence of the lipid architecture on the transition between the
lamellar and inverted-hexagonal phases is calculated, and qualitative agreement
with experiment is found.

\end{abstract}

\section{Introduction}
Lipid bilayers form the framework of biological membranes. Nevertheless
almost all membrane lipids adopt {\em non-bilayer} configurations, either in
their pure state or in lipid mixtures, under conditions close to
physiological ones \cite{luzatti,CULLIS}. Although cubic phases are also 
found \cite{luzatti,SEDDON3}, 
the most commonly occurring non-bilayer
arrangement is the inverted-hexagonal\cite{SEDDON}, or $H_{II}$, 
phase in which a matrix of
hydrocarbon tails is pierced by a hexagonal array of water-filled
tubes lined by the hydrophilic head groups. Why biological membranes
should contain lipids which tend to drive it towards a configurational
instability, much like a lamellar-hexagonal transition, 
has been the subject of much speculation centering on
the role such instabilities might play in promoting membrane fusion, and
controlling membrane permeability\cite{CULLIS,SIEGEL}. As a consequence
of this interest, there have been many studies of the phase behavior of
lipids in general, and of the parameters which affect the lamellar to
hexagonal transition in
particular\cite{SEDDONCEVC,GRUNER1,GRUNER,BRIGGS1,BRIGGS2}. For 
example, in
a homologous series of saturated diacyl or diakyl
phosphatidylethanolamines, an increase of chain length stabilizes the
$H_{II}$ phase,  causing the transition temperature between it
and the lamellar
phase, which exists at lower temperatures, to
decrease\cite{SEDDONCEVC}.
Conversely an increase in the volume of the head group 
stabilizes the lamellar phase, $L_{\alpha}$, causing the transition
temperature  to increase, as clearly 
demonstrated\cite{GRUNER} in mixtures of 
dioleoylphosphatidylethanolamine, DOPE, and
dioleoylphosphatidylcholine, DOPC. An increase in water
content also tends to stabilize $H_{II}$.

A qualitative understanding of these results is provided by a
characterization of the lipid as a simple geometrical object
parameterized by the chain volume, the maximum chain length, and the
head-group area \cite{Israel}.  The different phases result from simple
geometrical packing considerations. In contrast to simplifying the
description of the lipid, 
large-length scale approaches\cite{GRUNER,BILAYER} simplify the
description of the bilayer itself, reducing it to an infinitesimally
thin membrane characterized by elastic constants which are inputs to the
theories\cite{MINIMALSURFACES}. Often, elements of such theories are
combined with other phenomenological terms to complete the
description\cite{GRUNER}.

Simulation of chemically realistic models of lipids and their
interactions have been carried out\cite{SIMULATION}, and yield valuable
information about local properties, such as density and
orientation profiles in the bilayer, lamellar phase, and  dynamic
and transport properties. However simulations are
both demanding computationally and limited to a rather small number of
particles. Within the framework of  chemically realistic models, the
simulation of phase transitions between different morphologies seems not
to be feasible at present.

Analytic, mean-field, approaches combined with microscopic modeling of
the tails of lipids, have been applied with success to the manner in
which the tails pack in the interior of aggregates
\cite{Enumerate,Gruen,Szleifer}. The contribution of the chains to the
single-lipid partition function, required by mean-field theory, is
obtained from an enumeration of the molecular conformations of the tails
permitted within the Rotational Isomeric State (RIS)
model\cite{RIS}. It is here that the particular architecture of the
chains enters.  
Bilayer thickness is set by assumptions on a phenomenological free-energy 
describing the head-group region.
These calculations reproduce many features of the density
profiles and segment orientations in the interior of aggregates. 

The contribution of the lipid tails to the $H_{II}$ phase has also been
examined by these means\cite{STEEN}, and it was shown that their entropy always favors
this phase over the lamellar one. It was also observed that a change in
the area per head group could lead to a transition to the lamellar phase.
No solvent was included explicitly, but as its effect would be to alter the
area per head group, the observation indicates that variation of solvent
concentration would be able to bring about a transition.
The calculation is an approximate one by necessity because it is carried
out in real space. The full $p6mm$ symmetry of the $H_{II}$ phase
is not preserved, as the tubes are considered to be cylinders on which
the area per head group is uniform. Further, the packing constraints in
the interstices between the tubes cannot be satisfied exactly. 
Application of this real-space approach to such
complicated morphologies as $Ia\bar{3}d$ would be extremely 
laborious\cite{MILNER}.

In this paper, we overcome the difficulties of a real-space approach by
combining the above enumeration techniques with recent advances in the
solution of mean-field equations of polymers \cite{MATSENSCHICK,MATSEN},
a combination which we had previously shown to be fruitful in a
relatively simple system\cite{ROD}. Here we apply it to a more
complicated system, a minimal model of a lipid and solvent mixture, one
which treats the head group on the same microscopic basis as the tails.
We are able, thereby, to obtain its phase diagram within mean-field
theory and to examine how the boundary between $L_{\alpha}$ and $H_{II}$
phases changes with lipid architecture, {\em i.e.} how that architecture
affects the stability of the lamellar phase. Our results are in
qualitative agreement with experiment.

\section{Model and numerical Self-Consistent Field technique}
We describe the lipid as consisting of a
tail, comprised of any number of chains, 
(usually two), containing
altogether a total of $N$ identical segments of volume $v_t$ each, 
and a head containing two segments of volume $v_h/2$ each. The solvent
molecules have volume $v_s$ but are otherwise without structure.
The partition function of the mixture of $N_L$ lipid molecules and
$N_S$ solvent particles in  volume $V$ can be written

\begin{equation} {\cal Z} \sim \frac{1}{N_S!N_L!} \int 
 \prod_{\alpha=1}^{N_L}{\cal D}{\bf r}_{\alpha}  P[{\bf r}_{\alpha}] \;\;\; 
 \prod_{\beta=1}^{N_S}d{\bf r}_{\beta}
 \delta \left( 1-\hat{\Phi}_s - \frac{v_h}{v_s} \hat{\Phi}_h - 
   \frac{v_t}{v_s}\hat{\Phi}_t \right)
              \exp \left\{ \frac{-{\cal E}}{k_BT}\right\}
\end{equation}
where  
\begin{equation}
\hat{\Phi}_h = v_s \sum_{\alpha=1}^{N_L} \frac{1}{2} \sum_{h=1}^2\delta ({\bf
  r}-{\bf r}_{\alpha,h}), 
\end{equation}
\begin{equation}
\hat{\Phi}_t = v_s\sum_{\alpha=1}^{N_L} \sum_{t=1}^N \delta ({\bf r}-
{\bf r}_{\alpha,t}),
\end{equation}
are the dimensionless number densities of the head and the tail
segments, and 
\begin{equation}
\hat{\Phi}_s = v_s \sum_{\alpha=1}^{N_S} \delta ({\bf
  r}-{\bf r}_{\alpha,s}), 
\end{equation}
is the dimensionless number density of
solvent particles. The fluid has been treated as incompressible. The
probability distribution of the lipid configurations is denoted $P[{\bf
r}_{\alpha}]$, and the interaction energy between particles is ${\cal
E}$.

It is convenient to introduce auxiliary fields and
consider the particles to interact with one another via intermediate,
fluctuating fields rather than directly;
\begin{equation}
\label{pf}
{\cal Z} \sim \int {\cal D}\Phi_h {\cal D}\Phi_t {\cal D}\Phi_s 
                   {\cal D}W_h {\cal D}W_t {\cal D}W_s {\cal D}\Pi
                   \exp \left \{ - \frac{{\cal F}}{k_B T}\right \}
\end{equation}
where the free energy functional, ${\cal F}$, is 
\begin{eqnarray}
\frac{{\cal F }[\Phi_h,\Phi_t,\Phi_s,W_h,W_t,W_s,\Pi]}{ k_B
  T}  
 & = &
       \frac{{\cal E}}{  k_B T} - N_S \ln 
    \frac{{\cal Q}_S}{N_S} - N_L
       \ln \frac{{\cal Q}_L}{N_L} \nonumber \\
 - \frac{1}{v_s}\int d{\bf r}\; {\Big\{} W_h \Phi_h + W_t \Phi_t + W_s \Phi_s 
&&          + \Pi \left(1- \Phi_s - \frac{v_h}{v_s} \Phi_h 
     - \frac{v_t}{v_s} \Phi_t \right){\Big \}}
\end{eqnarray}

In this expression, ${\cal Q}_L$ denotes the single lipid partition
function in the external 
fields $W_h$ and $W_t$, acting on head
and tail segments, and ${\cal Q}_S$ the solvent partition function; 
\begin{equation}
{\cal Q}_L[W_h,W_t] = \int {\cal D}{\bf r}  P[{\bf r}] 
\exp \left \{ - \frac{1}{2}\sum_{h=1}^2 W_h({\bf r}_h) - 
\sum_{t=1}^{N} W_t({\bf r}_t)\right\},
\end{equation}
\begin{equation}
{\cal Q}_S[W_s] = \int d{\bf r}\; \exp {\Big \{} 
- W_s({\bf r}){ \Big \}}
\end{equation}

The interaction energy is
\begin{equation}
\label{initialint}
\frac{{\cal E}}{k_BT}=\frac{1}{2}\int\frac{d{\bf r}}{v_s}\frac{d{\bf
    r}'}{v_s}\sum_{\alpha,\beta}V_{\alpha\beta}({\bf r}-{\bf r}')
   \Phi_{\alpha}({\bf r})\Phi_{\beta}({\bf r}'),
\end{equation}
 where $\alpha$ and $\beta$ take the values $h$, $t$, and $s$. The
 incompressibility condition
\begin{equation}
\Phi_s({\bf r})+\frac{v_h}{v_s}\Phi_h({\bf
  r})+\frac{v_t}{v_s}\Phi_t({\bf r})=1,
\end{equation}
can be used to eliminate the solvent volume fraction $\Phi_s$ in
Eq.(\ref{initialint}). The terms linear in the head and tail volume
fractions which result from this procedure can be ignored, as 
they only contribute to the chemical
potentials of the heads and tails. 
At this stage we neglect the coupling between the molecular conformations, the local
fluid-like packing of molecular segments, and the local energy density and 
assume that all interactions are contact interactions, 
\begin{equation}
V_{\alpha\beta}({\bf r}-{\bf r}')=\epsilon_{\alpha\beta}\delta({\bf
  r}-{\bf r}').
\end{equation}
With these simplifications, the energy can be written in the form
\begin{equation}
\label{eq:int}
\frac{{\cal E}[\Phi_h,\Phi_t]}{k_B T} =\frac{1}{v_s} \int d{\bf r}\; 
\left\{ \chi_{ht} 
                    \Phi_h \Phi_t
                    -\frac{1}{2} \chi_{hh} \Phi_h^2
                    -\frac{1}{2} \chi_{tt} \Phi_t^2 \right\},
\end{equation}
with
\begin{eqnarray}
\label{chis1}\chi_{ht}& \equiv& \frac{1}{v_sk_BT}\left[\epsilon_{ht}
+\epsilon_{ss}\frac{v_hv_t}{v_s^2}-\epsilon_{hs}\frac{v_t}{v_s}-
\epsilon_{ts}\frac{v_h}{v_s}\right]\\
\chi_{hh}& \equiv&-\frac{1}{v_sk_BT}\left[\epsilon_{hh}+
\epsilon_{ss}\frac{v_h^2}{v_s^2}-2\epsilon_{hs}\frac{v_h}{v_s}\right]\\
\label{chis3}\chi_{tt}& \equiv& -\frac{1}{v_sk_BT}\left[\epsilon_{tt}+
\epsilon_{ss}\frac{v_t^2}{v_S^2}-2\epsilon_{ts}\frac{v_h}{v_s}\right
].
\end{eqnarray}

The functional integration in Eq. (\ref{pf}) cannot 
be carried out explicitly. Therefore we employ  mean-field theory, which
approximates the integral by the largest value of the integrand. 
This maximum occurs at values of the fields and densities which are determined 
by extremizing ${\cal F}$ with respect to each of its seven
arguments. Fluctuations around these most probable values are neglected.
These values are denoted below by lower case letters.
They satisfy the self-consistent equations
\begin{eqnarray}
\label{eq:mf1}
w_h({\bf r}) &=& \chi_{ht}\phi_t({\bf r})-\chi_{hh}\phi_h({\bf r})
 +\pi({\bf r}) v_h/v_s, \\
w_t({\bf r}) &=& \chi_{ht}\phi_h({\bf r})-\chi_{tt}\phi_t({\bf r}) 
+\pi({\bf r}) v_t/v_s,  \\
w_s({\bf r}) &=& \pi({\bf r}), \\
1 &=& \phi_s({\bf r})+\phi_h({\bf r}) v_h/v_s+\phi_t({\bf r}) v_t/v_s, \\
\phi_h({\bf r}) &=& - \frac{N_L v_s}{ {\cal Q}_L} 
\frac{\delta {\cal Q}_L}{\delta
  w_h({\bf r})}, \\
\label{phit}
\phi_t({\bf r}) &=& - \frac{N_L v_s}{ {\cal Q}_L} \frac{\delta {\cal Q}_L}{\delta
  w_t({\bf r})}, \\
\phi_s({\bf r}) &=& - \frac{N_S v_s}{{\cal Q}_S} \frac{\delta {\cal Q}_S}{\delta 
w_s({\bf r})}. 
\end{eqnarray}
Because the overall density
is fixed, we can set $\int d{\bf r}\pi({\bf r})$ $=\int d{\bf r}w_s({\bf
  r})= 0$. 
The mean-field free energy 
$F\equiv {\cal F}[\phi_h,\phi_y,\phi_s,w_h,w_t,w_s,\pi]$ 
is 
\begin{eqnarray}
&&\frac{v_s F}{ V k_B T}= \phi_{s} \ln \phi_{s} + \phi_{L} \ln \phi_{L}
                           - \phi_{s} \ln \frac{{\cal Q}_S v_s}{V} 
                          - \phi_{L} \ln \frac{{\cal Q}_L v_s}{V}  
\nonumber \\
&&                          
            + \frac{1}{V} \int d{\bf r}\; \left \{
                            - \chi_{ht} \phi_h({\bf r}) \phi_t({\bf r})
                             +\frac{1}{2} \chi_{hh} \phi_h^2({\bf r}) + 
\frac{1}{2} \chi_{tt} \phi_t^2({\bf r})\right \},
\end{eqnarray}
where we have denoted $\phi_{s}\equiv N_sv_s/V$, the average,
dimensionless, number density of solvent, and similarly
$\phi_{L}\equiv N_Lv_s/V$, for lipids.
 
In order to study the self-assembly of lipids into various morphologies,
we expand the spatial dependence of the densities and fields in a
complete set of orthonormal functions $\{ f_k({\bf r}) \}$, $V^{-1} \int
d{\bf r}\; f_if_k = \delta_{ik}, \; f_1=1$, which possess the symmetry
of the morphology being considered\cite{MATSENSCHICK}; {\em e.g.}
$\phi_h({\bf r}) = \sum_k \phi_{h,k} f_k({\bf r})$. 
The coefficients $\phi_{h,1}$, $\phi_{t,1}$, and $\phi_{s,1}$ are simply
equal to the average, dimensionless number densities $\phi_h$, $\phi_t$, and 
$\phi_s$ respectively. 
The solvent
density $\phi_s({\bf r}) = \phi_{s,1} V\exp[-w_s({\bf r})]/\int d{\bf r}\; 
\exp[-w_s]({\bf r})$
can be Fourier expanded as 
$\phi_{s,k} = \phi_{s,1} (\exp S)_{k,1}/(\exp S)_{1,1},$
with 
\begin{equation}
\qquad (S)_{k,m} = -\sum_n w_{s,n} \frac{1}{V} \int d{\bf r}\; f_k f_n f_m.
\label{eq:mf72}
\end{equation}

The partition function of a single lipid in an external field cannot
be obtained analytically for a realistic architecture. Therefore we
approximate the non-interacting single lipid probability distribution 
${\cal P}$ by a
representative sample of ${\cal N}$ single lipid conformations. 
Assigning the Boltzmann weight, $\omega_c$, to each lipid conformation 
in the field of mean potentials $w_h$ and $w_t$
\begin{equation}
\omega_c \equiv \exp \left \{  - \sum_k \left( w_{h,k} \frac{1}{2}\sum_{h=1}^2 f_k({\bf r}_{c,h}) 
                       + w_{t,k} \sum_{t=1}^{N} f_k({\bf r}_{c,t})
                     \right) \right\}
\end{equation}
we obtain from Eq.\ (\ref{phit}) 
the components of the tail segment density 
\begin{equation} 
\label{eq:mf66}
\phi_{t,k} = \phi_{t,1} \frac{\sum_{c=1}^{\cal N} 
\omega_c \sum_{t=1}^N f_k({\bf r}_{c,t})}{\sum_{c=1}^{\cal N} \omega_c}, 
\end{equation}
and similarly for the head segment density.
The self-consistent equations, expressed in the basis $\{f_k\}$, are solved
by a Newton-Raphson-like method.  Finally we minimize the free
energy with respect to the size of the unit cell, $D$. To this end we
translate the lipid conformations so as to achieve a uniform distribution in
the cell.

\section{Results}
\begin{figure}[htbp]
  \begin{minipage}[t]{160mm}%
    \setlength{\epsfxsize}{13cm}
       \mbox{\epsffile{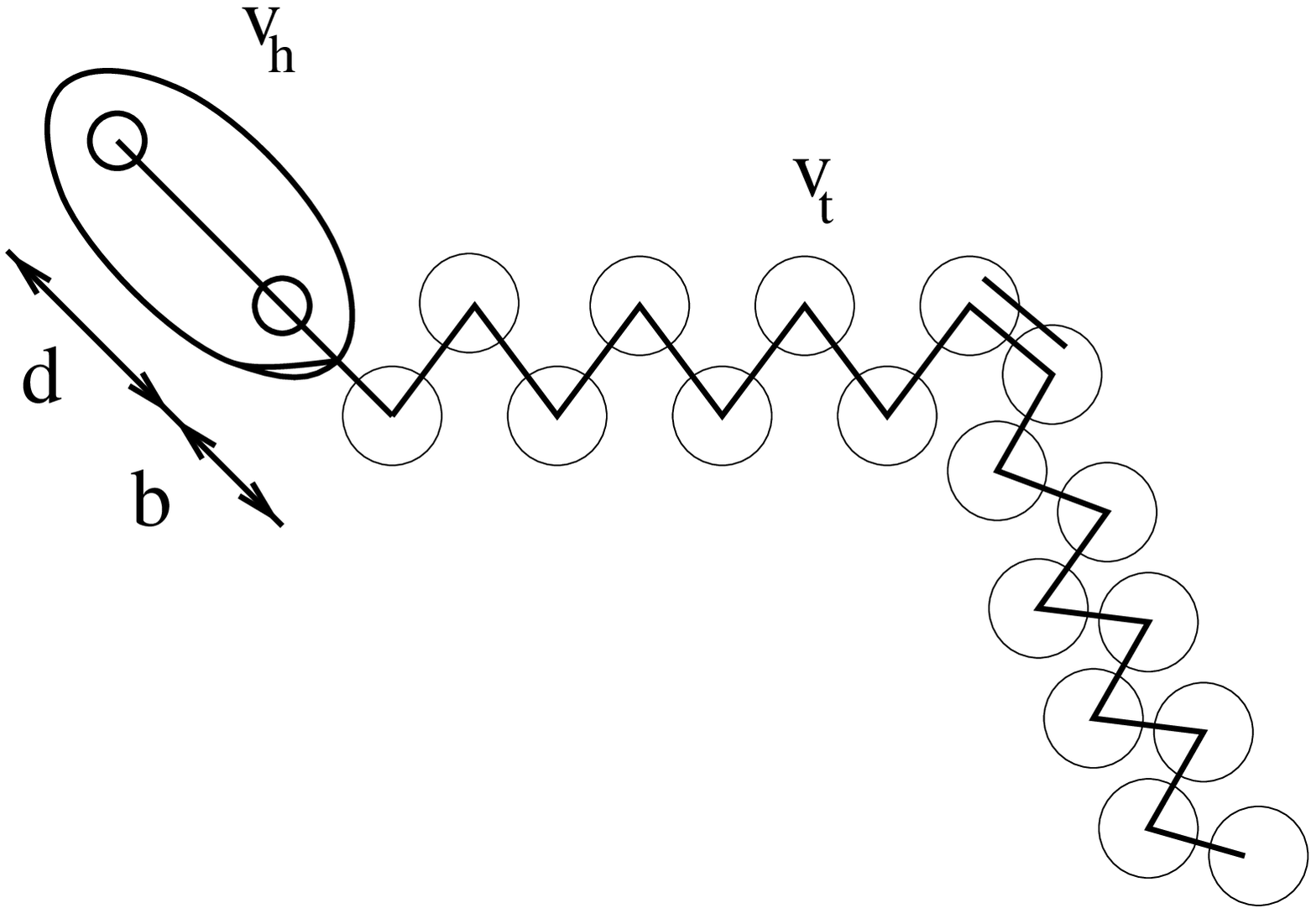}}
  \end{minipage}%
  \hfill%
  \begin{minipage}[b]{160mm}%
    \caption{Schematic drawing of the lipid architecture of our model.
             The specific volumes of the head and of a 
             single tail segment are $v_h$ and $v_t$ respectively.
            The distance between the tail segments is $b$, and that
            between the two segments of the head is $d$.
      }
    \label{fig:picture}
  \end{minipage}%
\end{figure}

The above scheme is applicable to {\em arbitrary} lipid architecture and
symmetry
of spatial ordering. We have applied it to model monoolein, a fatty acid whose 
phase
behavior in water has attracted much interest\cite{BRIGGS2,OLEIN}. 
A schematic sketch of the architecture is presented in Fig.\ \ref{fig:picture}
to illustrate the architectural parameters of our model.
The single hydrocarbon tail contains $N=17$ units, a distance $\ell=1.53$\AA\  
apart, with a double bond between the 8th and 9th units. The volume of the 
tail units is $v_t=29$ \AA$^3$.
The trans-gauche energy difference is taken to be 500cal/mole. We take the two
units of the head and the first segment of the tail to be collinear, with a
distance $d$ between the head units, and a distance $b$ between the
first tail segment and the adjacent head unit. 
We set all interactions to zero save those between hydrophilic and 
hydrophobic entities, and they are taken to have the same strength
$\epsilon$.
Thus from Eqs. (\ref{chis1}-\ref{chis3}), $\chi_{hh}=0$, 
$\chi_{tt}=(2\epsilon v_t/k_BTv_s^2)\equiv\chi$, and 
$\chi_{ht}=\chi(v_s-v_h)/(2v_t)$.
Typically between 524,288 and 2,097,152 conformations of single lipids
were generated.  The position of the head was chosen uniformly
distributed over the unit cell, with a random orientation. The tail was
then constructed according to the Rotational Isomeric State (RIS)
model\cite{RIS}. The tail conformations were generated at a temperature
of 300K, so that they are fluid.  

We use up to 32 basis functions, $f_k({\bf r})$.  Because the
calculations involve the contributions of each individual lipid to the
Fourier component of the density, the required computer memory scales
like the product of lipid conformations and number of basis
functions. The program needs more than 1 Gbyte memory.  We employ a
massively parallel computer CRAY T3D/T3E and distribute the lipid
conformations among the processors independently of their spatial
position.  Each processor evaluates the contribution of its assigned
lipid conformations to the Fourier components of the density according
to eqn. (\ref{eq:mf66}).  The partial results are collected via {\tt
shmem} routines. The first processor also calculates the solvent
density, and therefore we assign it a smaller number of lipid
conformations to compensate for the additional work load.  Typically
between 32 and 128 processors have been employed in parallel. The
program scales linearly with the number of processors.

\begin{figure}[htbp]
  \begin{minipage}[t]{160mm}%
    \setlength{\epsfxsize}{13cm}
       \mbox{\epsffile{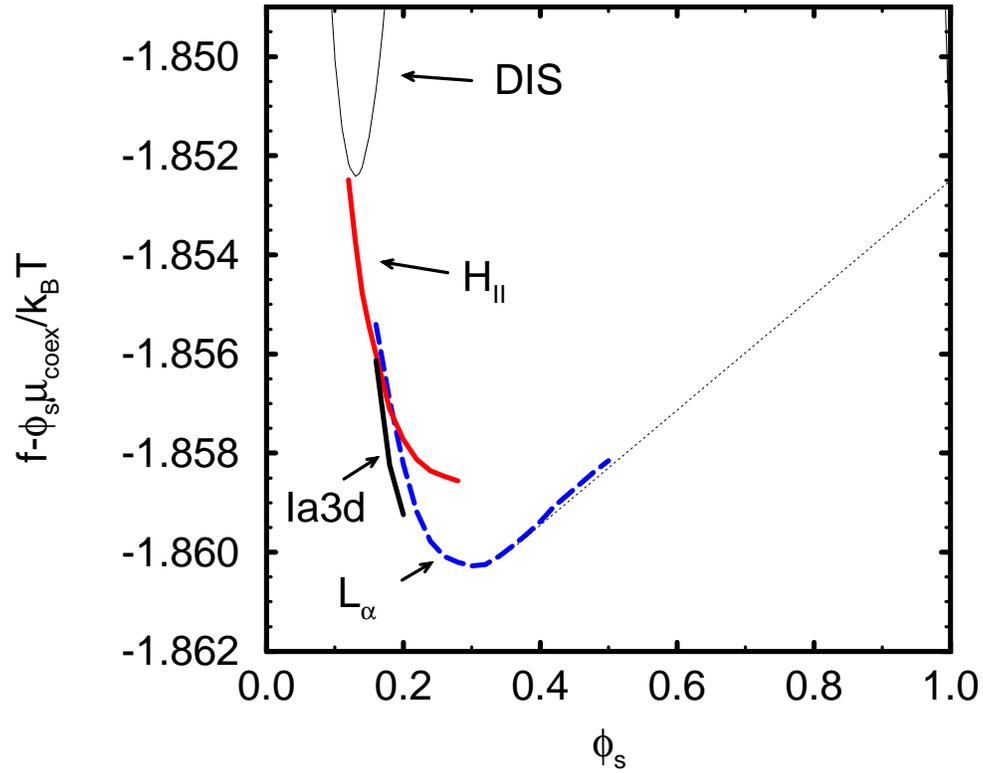}}
  \end{minipage}%
  \hfill%
  \begin{minipage}[b]{160mm}%
    \caption{Free energies at $1/\chi_{tt}=0.25$ of the disordered,
      $L_{\alpha}$, $Ia\bar{3}d$, and $H_{II}$ phases
      as function of the solvent content. For convenience, 
      the term $\mu_{coex}\phi_s/k_BT$ has been subtracted from the
      dimensionless free-energy density $f$, where $\mu_{coex}$ is the
      chemical potential at which solvent-rich and solvent-poor
      disordered phases would coexist were there no stable ordered
      phases. The dotted line shows the Maxwell construction between
      the $L_{\alpha}$ and the solvent-rich disordered phase.
      }
    \label{fig:free_energy1}
  \end{minipage}%
\end{figure}

We first consider the parameters $v_t/v_s=1$, $v_h/v_s=3.2$, $d/\ell=0$,
and $b/\ell=1$. Thus the head is a relatively small, single, interaction
center located the same distance from the first tail segment as each
tail segment is from its neighbors on the chain. The ratio of 
solvent volume, $v_s$, to lipid volume, $v_l=v_h+N v_t$, 
determines the extent to
which the microstructure can be swollen. Small solvents tend to swell
the microstructure due to their large translational entropy, whereas
large solvents favor phase separation. The phase diagram is determined
by calculating the free energy of the different morphologies as a
function of the solvent concentration and the ``temperature''
$1/\chi$. Phase coexistence is determined by equating the chemical
potential

\begin{equation}
\mu=\frac{\partial F}{\partial N_s}=\frac{v_s\partial F}{V\partial
 \phi_{s}},
\end{equation}
and the Gibbs free energy
\begin{equation}
\frac{G}{V}=\frac{F}{V}-\frac{\mu\phi_{s}}{v_s},
\end{equation}
in the two phases.
For a representative
value of the temperature, $1/\chi=0.25$, we find the sequence of
dimensionless
free
energy densities, $f\equiv v_sF/Vk_BT$, shown in Fig.\ \ref{fig:free_energy1}. 
It follows from this sequence that, with
increasing solvent concentration, there is a transition from the
disordered phase to the $H_{II}$ phase, and from that to the $Ia\bar{3}d$
cubic phase. At larger concentrations, the lamellar $L_{\alpha}$ phase 
competes with coexistence between water-rich and $Ia\bar{3}d$ phases.
This is in accord with the phase diagram of 
monoolein\cite{BRIGGS2,OLEIN}.

Having determined that the $Ia\bar{3}d$ phase does occur in our
calculation, we do not consider it further. A large number of basis
functions is required to determine its free energy with sufficient
accuracy to determine its phase boundaries. Further, it is sufficient to
restrict ourselves to the inverted-hexagonal and lamellar morphologies
to investigate the effect of lipid architecture on their relative
stability, which is our principal interest here.

\begin{figure}[htbp]
    \begin{minipage}[t]{160mm}%
       \setlength{\epsfxsize}{13cm}
       \mbox{\epsffile{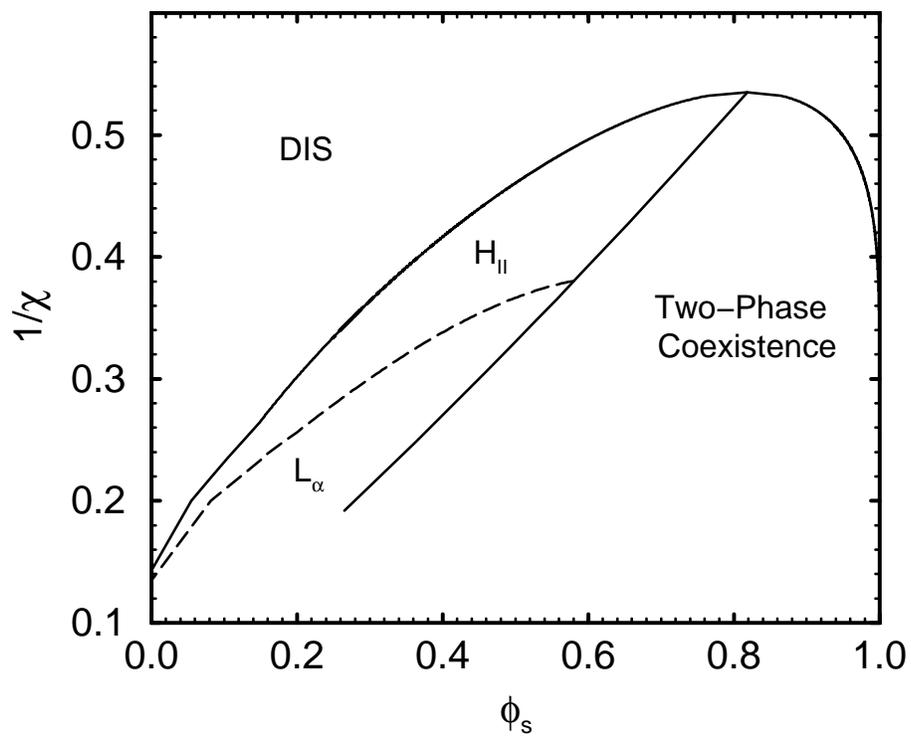}}
    \end{minipage}%
    \hfill%
    \begin{minipage}[b]{160mm}%
       \caption{Phase diagram of the model with $v_t/v_s=1$,
         $v_h/v_s=3.2$, 
         $d/{\ell}=0$, $b/{\ell}=1$.
                The stability region of cubic phases is not included.
                }
       \label{fig:phase1}
    \end{minipage}%
\end{figure}

The calculated phase diagram of the system, excluding the cubic phases,
is shown in Fig.\ \ref{fig:phase1}.  Its salient features are similar 
to those of the experimental monoolein/water
mixture\cite{BRIGGS2,OLEIN}, and other lipid, solvent
mixtures\cite{cevc}. The $H_{II}$ phase tends to exist at higher temperatures
and water concentrations than does the $L_{\alpha}$ phase.
Upon swelling the $H_{II}$ with solvent at lower
temperatures, 
a weak first-order transition to another lyotropic phase, ($L_{\alpha}$
in this calculation), is encountered, 
while at higher temperatures, a strong first-order transition to a disordered, 
(DIS), solvent-rich phase, is seen. These features are all in agreement with 
experiment.

\begin{figure}[htbp]
    \begin{minipage}[t]{160mm}%
       \setlength{\epsfxsize}{9cm}
       \mbox{\epsffile{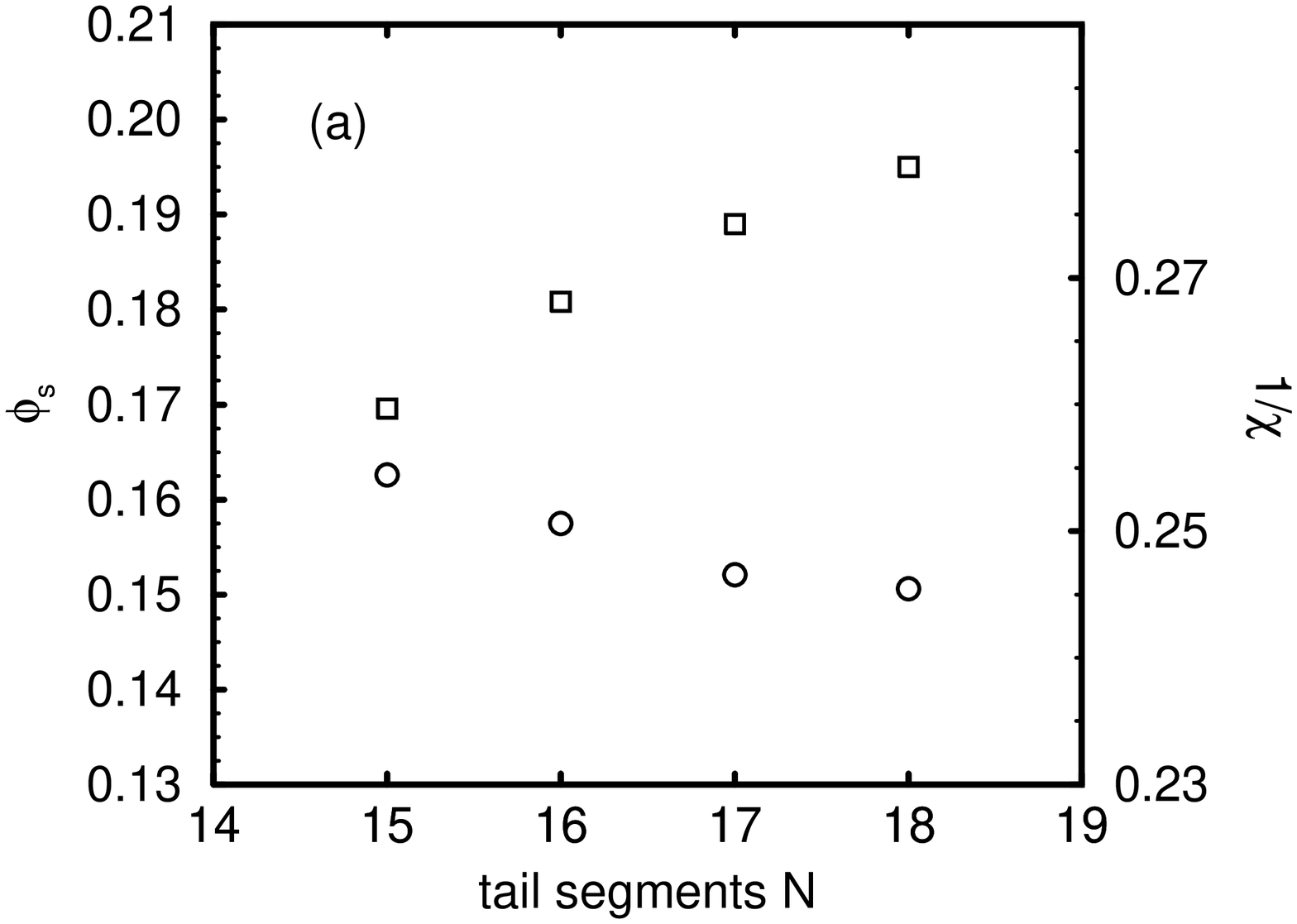}}
       \setlength{\epsfxsize}{9cm}
       \mbox{\epsffile{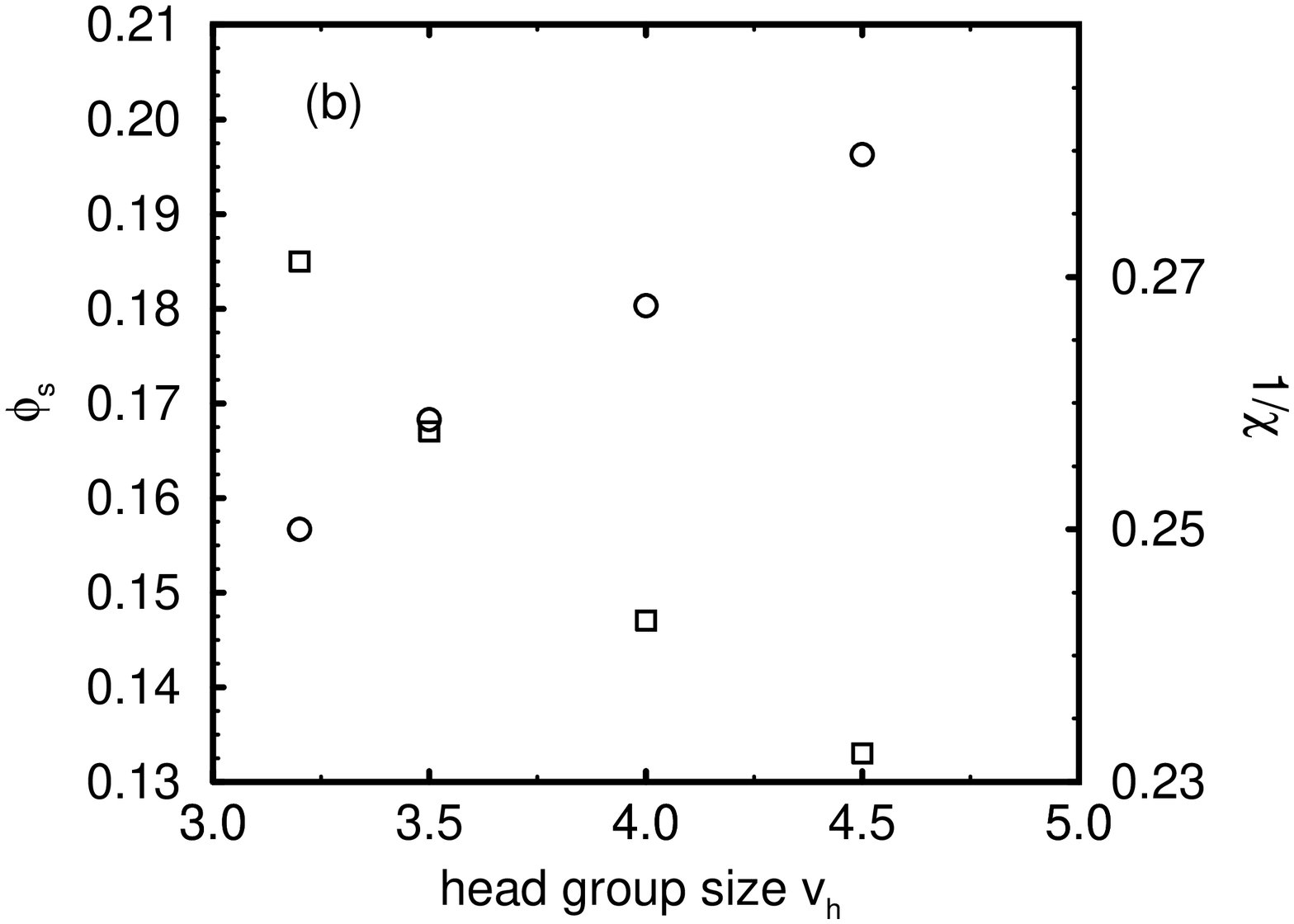}}
    \end{minipage}%
    \hfill%
    \begin{minipage}[b]{160mm}%
       \caption{Influence of lipid architecture on the lamellar to inverted
                hexagonal transition. The squares denote the 
                density $\phi_s$ along the coexistence curve at fixed
                $1/\chi=0.25$,
                whereas the circles represent the transition temperature,
                $1/\chi$, at fixed composition $\phi_s=0.185.$
                {\bf (a)} Variation with the chain length $N$. 
                {\bf (b)} Variation with the head group size $v_h$.
                }
       \label{fig:arch1}
    \end{minipage}%
\end{figure}

The effect of lipid architecture on the $L_{\alpha}$, $H_{II}$,
transition is shown in Fig.\ \ref{fig:arch1}.  In (a), we see that 
at fixed temperature $1/\chi=0.25$, the
solvent concentration within the narrow coexistence region 
between $H_{II}$ and $L_{\alpha}$ phases,
shown by squares, increases with
increasing tail length, while the temperature on the phase boundary at
fixed concentration, $\phi_s=0.185$, shown by circles, decreases. Thus
lengthening the tail stabilizes the $H_{II}$ phase, in agreement with
experiment\cite{SEDDON}. The tails in this calculation were taken to be
fully saturated so that there would be no effect of the relative
placement of the double bond.  The effect of changing the head group
volume is shown in (b), and it is seen that increasing the volume of the
head group destabilizes the $H_{II}$ phase, again in agreement with
experiment\cite{GRUNER}.  

A posteriori, this behavior can also be rationalized in the framework of
packing models. The
morphology is controlled by a packing parameter $\eta = v_l/al_c$, where
$a$ denotes the area per head group and $l_c$ the maximum extension of
the lipid tail.  We set $l_c = \sqrt{\langle R^2_{ht}\rangle} = 15.1 \AA$, where
$\langle R^2_{ht}\rangle$ is the ensemble average of the square of the distance 
between the head unit and the
last tail segment. The area per lipid tail can be related to the repeat
distance $D$ via simple geometric considerations:
\begin{equation}
a_{\rm lam} = \frac{2 (v_h+Nv_t)}{D (1-\phi_{s,1})} 
\qquad \mbox{lamellar phase}
\end{equation}
\begin{equation}
 a_{\rm hex} = 
\frac{2  (v_h+Nv_t)}{D (1-\phi_{s,1})} \sqrt{\frac{2\pi}{\sqrt{3} }
\left( \phi_s+\frac{(1-\phi_s)v_h}{2(v_h+Nv_t)}\right)}
\qquad \mbox{inverted-hexagonal phase}
\end{equation}
There is a transition from an inverted-hexagonal to a lamellar phase
upon decreasing the packing ratio $\eta$. The molecule is pictured as a
wedge with the tails constituting the bulkier part. An increase of the
head group volume or decrease of the hydrocarbon tail length reduces the
effective wedge shape of the molecule, and therefore tends to stabilize
the lamellar phase.

\begin{figure}[htbp]
    \begin{minipage}[t]{160mm}%
       \setlength{\epsfxsize}{13cm}
       \mbox{\epsffile{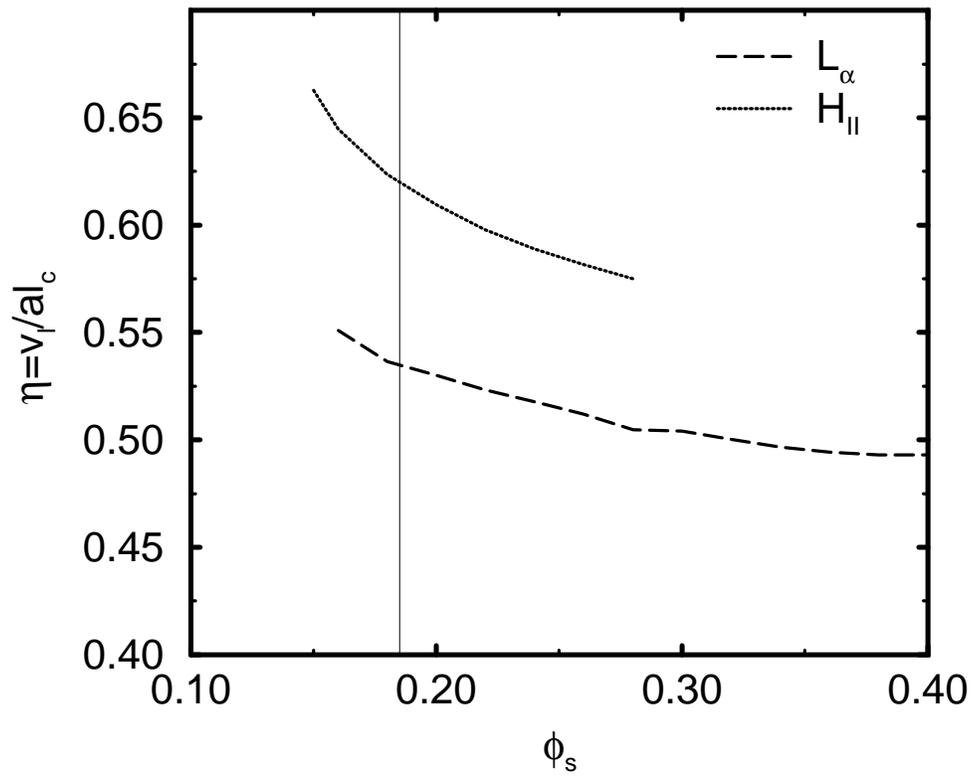}}
    \end{minipage}%
    \hfill%
    \begin{minipage}[b]{160mm}%
       \caption{Packing parameter for the lamellar $L_\alpha$ and 
                inverted-hexagonal $H_{II}$ phase 
                as a function of the solvent concentration for $1/\chi=0.25$.
                The vertical line marks 
                solvent composition at which the transition occurs.
                }
       \label{fig:israel}
    \end{minipage}%
\end{figure}

The occurrence of a transition upon increase of the solvent content is
understood as due to the swelling of the area per head group, which also
reduces the effective wedge shape of the molecule.  The results of our
self-consistent field calculations are displayed in Fig.\
\ref{fig:israel}.  Calculating the packing parameter from a microscopic
model, we find that it decreases upon adding solvent, and takes a value
between $0.5$ and $0.65$ at the transition from the inverted-hexagonal to the
lamellar phase. From the simple packing arguments, one would have
expected its value to exceed unity at the transition. This discrepancy
only illustrates that the phenomenological parameters of the packing model
are  related but approximately to the geometrical parameters of the
molecule.

Along the $H_{II}$, $L_{\alpha}$ transition, we find that the ratio of
the lattice constants of the two coexisting phases is $D_H/D_L\approx
1.10$, rather close to the value of 1.16 extrapolated from the
experiments on monoolein\cite{BRIGGS2}. However the absolute values of
the lattice spacings are smaller than the experimental ones. The latter
are $D_L=$ 42\AA\ while we obtain 20\AA.  This small value 
implies that the head groups are separated by a very
thin water layer, and that hydrocarbon tails originating from
different monolayers interdigitate significantly, a shortcoming
encountered in other calculations\cite{LEERMAKERS}.  By making the head
group bulkier, $d/\ell=6$, and adjusting the head group volume
$v_h/v_s=8.5$ such that the transition occurs at the same solvent
density, we increase the calculated result to $D_L$=28\AA.  Extending
the head group still further, to take account of their hydration shell,
would improve the agreement with experiments.

\section{Discussion}
We have explored the self-assembly of monoacyl lipids in solution employing
a microscopic model. The model describes the architecture of the tail
very well, by means of  the Rotational Isomeric State scheme, and that
of the head rather crudely. We have 
calculated the phase diagram employing no other assumption than that of 
 mean field theroy. In agreement with experiments, we find that
a transition from
an inverted-hexagonal to a lamellar phase occurs  
upon increasing the
solvent content or decreasing the temperature, and that
an increase in the length of the hydrocarbon tail or
a decrease in the head group volume stabilizes the inverted-hexagonal phase.
The ratio of the lattice constants of the coexisting phases is in reasonable
agreement with experiment. However the absolute value of the lamellar
spacing is too low.
This result highlights one of two limitations of our calculation; that
we have greatly simplified the head group and the solvent.  We expect
that this deficiency can be remedied with more realistic
parameterization of the interactions.  We have also ignored
fluctuations, which are expected to shift the phase boundaries to
somewhat lower temperatures and, by inducing effective repulsions
between interfaces, to enlarge the characteristic length scale of the
phases.  Because of the extended molecular architecture, these effects
will be small.  They will also have little effect on the stability of
the lyotropic phases relative to one another.  Therefore we are hopeful
that our approach, having demonstrated its utility in capturing the
effect of architecture on lipid phase transitions, will be applicable to
the more difficult problem of its role in membrane fusion and
permeability.

\subsection*{Acknowledgment}
It is a pleasure to thank W.  Frey, J.  Seddon, I.  Szleifer, and P.  Yager 
for stimulating conversations.  Financial support by the Alexander von 
Humboldt foundation, and the National Science Foundation under Grant No.\ 
DMR9531161, as well as CRAY T3D/T3E time at the San Diego Supercomputer 
Center are gratefully acknowledged.

\end{document}